\documentstyle[epsfig]{mn2e}

\def\be{\begin{equation}}
\def\ee{\end{equation}}
\def\bea{\begin{eqnarray}}
\def\eea{\end{eqnarray}}

\title[]{The X-ray afterglow flat segment in short GRB 051221A:
Energy injection from a millisecond magnetar?}
\author[]{Yi-Zhong Fan$^{1,2,3}$\thanks{Lady Davis Fellow, E-mail: yzfan@pmo.ac.cn},
and Dong Xu$^4$\\
$^1${\sl The Racah Inst. of Physics, Hebrew University, Jerusalem 91904, Israel}\\
$^2$ {\sl Purple Mountain Observatory, Chinese Academy of
Science, Nanjing 210008, China}\\
$^3${\sl National Astronomical Observatories, Chinese Academy of
Sciences, Beijing 100012, China}\\
$^4${\sl Dark Cosmology Centre, Niels Bohr Institute, University
of Copenhagen, Juliane Maries Vej 30, 2100 Copenhagen, Denmark}\\}
\date{Accepted ......
Received ......; in original form ......}

\begin{document}

\maketitle
\begin{abstract}
The flat segment lasting $\sim 10^4$ seconds in the X-ray
afterglow of GRB051221A represents the first clear case of strong
energy injection in the external shock of a short GRB afterglow.
In this work, we show that a millisecond pulsar with dipole
magnetic field $\sim 10^{14}$ Gauss could well account for that
energy injection. The good quality X-ray flat segment thus
suggests that the central engine of this short burst may be a
millisecond magnetar.
\end{abstract}

\begin{keywords}
Gamma Rays: bursts$-$GRBs: individual (GRB 051221A)$-$ISM: jets
and outflows--radiation mechanisms: nonthermal
\end{keywords}

\section{Introduction}
\label{sec:Observation}

GRB 051221A was localized by the Burst Alert Telescope (BAT) onboard the {\it Swift}
satellite (Parsons et al. 2005) and promptly observed by both \emph{Swift}/BAT and
the Konus-Wind instrument. The Swift observations reveal this is a short hard burst,
with $T_{90}=1.4\pm 0.2$ s, a hard photon index $\alpha=-1.39\pm 0.06$, and a
fleunce $1.16\pm0.04 \times 10^{-6}\,\rm{ergs\,cm^{-2}}$ in the 15-150 KeV band
(Cummings et al. 2005). The Konus-Wind cutoff power-law spectral fitting, in the
20-2000 KeV band, shows a fluence $3.2^{+0.1}_{-1.7}\times 10^{-6}~{\rm
erg~cm^{-2}}$, a low-energy photon index $\alpha=-1.08\pm 0.14$, and an observed
peak energy $E_{\rm peak}=402^{+93}_{-72}$ KeV (Golenetskii et al. 2005). With a
redshift $z=0.5459$ (Soderberg et al. 2006), this burst's isotropic prompt emission
energy is $E_\gamma \sim 2.4\times 10^{51}$ erg, using the $\Lambda$CDM concordance
model of $\Omega_M=0.27$, $\Omega_\Lambda=0.73$, and $h=0.71$.

Both the X-ray ($\sim 10^2 -2\times 10^6$ s) and the optical
($\sim 10^4-4\times 10^5$ s) afterglow light curves of GRB051221A
have been well detected, while in the radio band only one data
point followed by several upper limits is available. This burst is
distinguished by an X-ray flattening at $t\sim 0.03-0.2$ day,
which strongly suggests a significant energy injection (Soderberg
et al. 2006; Burrows et al. 2006). However, the nature of that
energy injection is not clear. In the widely accepted double
neutron star merger model for the short/hard burst, supported by
the lack of detection of the bright supernova component in the
current event (Soderberg et al. 2006), the material ejected in the
merger is $\sim (10^{-4}-10^{-2})M_\odot$ (Rosswog et al. 1999;
Ruffert \& Janka 2001). Given an energy conversion efficiency
$\sim 0.001$, the fall-back accretion of part of that material
onto the central post-merger object is not likely to be able to
pump energy up to $\sim 10^{52}$ erg, even with moderate beaming
correction. So the fall-back accretion model, which may give rise
to significant energy injection in the collapsar scenario of
long/soft GRBs (MacFadyen, Woosley \& Heger 2001), does not work
in the current case.

In this Letter, we'll show that the afterglow undergoing an energy
injection from a millisecond pulsar with a dipole magnetic field
$\sim 10^{14}$ Gauss (i.e., a magnetar) well accounts for the
multi-wavelength data. The good quality X-ray flat segment thus
suggests that the central engine of this short burst may be a
millisecond magnetar.

\section{Analytical investigation}
In the long/soft GRB scenario, the energy injection caused by a millisecond pulsar
has been discussed in some detail (Dai \& Lu 1998; Wang \& Dai 2001; Zhang \&
M\'esz\'aros 2001; Dai 2004; Ramirez-Ruiz 2004; Zhang et al. 2006). Similarly,
provided that the gravitational wave radiation is not important, the dipole
radiation luminosity of a magnetar can be estimated by
\begin{equation}
L_{\rm dip}(t_b)\simeq 2.6\times 10^{48}~{\rm
erg~s^{-1}}~B_{\bot,14}^2R_{s,6}^6\Omega_4^4(1+t_b/T_o)^{-2}, \label{eq:L_dip}
\end{equation}
where $B_{ \bot}$ is the dipole magnetic field strength of the magnetar, $R_s$ is
the radius of the magnetar, $\Omega$ is the initial angular frequency of radiation,
the subscript ``$b$" represents the time measured in the burst frame, $T_o=1.6\times
10^4 B_{\bot,14}^{-2} \Omega_4^{-2}I_{45}R_{s,6}^{-6}$ s is the initial spin-down
timescale of the magnetar, $I\sim 10^{45}~{\rm g~cm^2}$ is the typical moment of
inertia of the magnetar (Pacini 1967; Gunn \& Ostriker 1969). Here and throughout
this text, the convention $Q_x=Q/10^x$ has been adopted in cgs units.

The energy emitted at $t_b$ will be injected into the previous GRB ejecta at time
$T_b$ satisfying
\begin{equation}
\int^{T_b}_{t_b}[1-\beta(\tau_b)]c d\tau_b=c\int^{t_b}_0
\beta(\tau_b) d\tau_b\approx c t_b, \label{eq:time1}
\end{equation}
where $\beta$, in units of the speed of light $c$, is the velocity of the ejecta
moving toward us. The corresponding observer time is
\begin{equation}
t \approx (1+z)\int^{T_b}_0[1-\beta(\tau_b)]d\tau_b.
\label{eq:time2}
\end{equation}
Equations (\ref{eq:time1}) and (\ref{eq:time2}) yield $t \approx
(1+z)t_b+(1+z)\int^{t_b}_0[1-\beta(\tau_b)]d\tau_b \approx (1+z)t_b$.

At time $T_b$, the energy injected into the ejecta satisfies
$dE/dT_b=[1-\beta(T_b)]L_{\rm dip}(t_b)$. With the relation
$dt=(1+z)[1-\beta(T_b)]dT_b$, we have
\begin{eqnarray}
&&{dE \over dt} \approx {1\over 1+z} L_{\rm dip}({t\over
1+z})\nonumber\\
&&= {2.6\times 10^{48} \over (1+z)}~{\rm
erg~s^{-1}}~B_{\bot,14}^2R_{s,6}^6\Omega_4^4[1+{t\over (1+z)T_o}]^{-2}.
\label{eq:E_inj}
\end{eqnarray}
So the energy injection rate $dE/dt \sim {\rm const}$ for $t\ll (1+z)T_o$ and $dE/dt
\propto t^{-2}$ for $t\gg(1+z)T_o$.

A general energy injection form can be written as $dE/dt
=A(1+z)^{-1} (t/t_o)^{-q}$ for $t_{i}<t<t_{f}$, where $t_i$ and
$t_{f}$ are the times when the energy injection takes place and
turns off, respectively (Zhang \& M\'esz\'aros 2001; Zhang et al.
2006). GRB ejecta's dynamical evolution, at a time $t_c$, is
significantly changed when the injected energy roughly equals to
its initial kinetic energy, i.e., $\int_{t_i }^{t_c } {(dE/dt)} dt
\sim E_k$. We thus derive $At_o^{\rm q}(t_c^{\rm 1-q}-t_i^{\rm
1-q})\sim (1+z)(1-q)E_k$. Accordingly, our magnetar model requires
$t_o=1$, $t_i\sim 0$ and $q\sim 0$ for $t_c<(1+z)T_o$, which leads
to $A\sim (1+z)E_\gamma/t_c$, and so
\begin{equation}
2.6\times 10^{48}~{\rm erg~s^{-1}}~B_{\bot,14}^2R_{s,6}^6\Omega_4^4\sim
(1+z)E_k/t_c. \label{eq:relation1}
\end{equation}
For $t>(1+z)T_o$, the rate of the energy injection drops sharply
or even the central supermassive magnetar has collapsed when it
has lost significant part of the angular momentum, which indicates
that the afterglow lightcurve flattening weakens at the time
$t_f\geq (1+z)T_o$. So we have
\begin{equation}
1.6\times 10^4 (1+z)B_{\bot,14}^{-2} \Omega_4^{-2}I_{45}R_{s,6}^{-6} \sim t_f.
\label{eq:relation2}
\end{equation}
>From equations (5) and (6), the total injected energy could be estimated as $E_{\rm
inj}=t_f E_k/t_c\sim 5\times 10^{52}~{\rm erg}~I_{45} \Omega_4^2$. On the other
hand, with and without energy injection, the contrast of forward shock X-ray
emission flux can be estimated by (Kumar \& Piran 2000)
\begin{equation}
f \sim (E_{\rm inj}/E_k)^{\rm (p+2)/4}\sim [5\times 10^{52}~{\rm
erg}~I_{45} \Omega_4^2/E_k]^{\rm (p+2)/4},  \label{eq:relation3}
\end{equation}
where $p\sim 2.4$ is the power-law index of the shocked electrons. Equations
(\ref{eq:relation1})$-$(\ref{eq:relation3}) are our main relations to constrain the
physical parameters of the underlying magnetar.

From the X-ray observations of GRB051221A, we measure $t_c\sim
3000\,{\rm s}$, $t_f\sim 1.5\times 10^4\,{\rm s}$, and $f\sim 6$.
Substitute $p$ and $f$ into equation (7), we find $E_k \sim
7.6\times 10^{51}I_{45}\Omega_4^2\,{\rm erg}$ is well consistent
with the observational result $E_k\sim 2E_{\gamma} \sim 4.8\times
10^{51}\,{\rm erg}$ when taking typical $I_{45}\sim 1.5$,
$\Omega_4\sim 0.65$ (i.e. 1 millisecond period), and the GRB
efficiency $\eta=E_k/(E_k+E_\gamma)\sim 30\%$. Furthermore, the
measurements of $t_c$, $t_f$, $f$, and $p$ are well consistent
with the constraint relation $f\simeq(t_f /t_c)^{\rm (p+2)/4}$.
{\it So we conclude that the central engine may be a magnetar}.
Its physical parameters are $(\Omega, R_{s}, B_{\bot}, I)\sim
(6500\,{\rm s}^{-1}, 13\,{\rm km}, 10^{14}\,{\rm Gauss}, 1.5\times
10^{45}\,{\rm g~cm^2})$ according to the above constraint
relations.

An additional constraint is on the ellipticity $\epsilon$ of the
magnetar. As shown in Shapiro \& Teukolsky (1983), the spin-down
timescale due to the gravitational wave radiation is $\tau_{\rm
gw}\sim 3\times 10^{-3}\epsilon^{-2}I_{45}^{-1}\Omega_4^{-4}~{\rm
s}$. Now $\tau_{\rm gw}>t_f\sim 1.5\times 10^4~{\rm s}$, we have
$\epsilon < 5\times 10^{-4} I_{45}^{-1/2}\Omega_4^{-2}$.

\section{Numerical fit to the afterglows of GRB 051221A}
We interpret the apparent flattening in the X-ray light curve
being caused by an energy injection from the central magnetar. Yet
the flattening episode is unapparent in the optical and radio
bands because of limited observations. The code here to fit the
multi-band lightcurves has been used in Fan \& Piran (2006) and
Zhang et al. (2006), and has been confirmed by J. Dyks
independently (Dyks, Zhang \& Fan 2006). The dynamical evolution
of the outflow is calculated using the formulae in Huang et al.
(2000), which are able to describe the dynamical evolution of the
outflow for both the relativistic and the non-relativistic phases.
One modification is that now we have taken into account the energy
injection (for instance see equation (12) of Fan \& Piran (2006);
see also Wei, Yan \& Fan 2006). The electron energy distribution
is calculated by solving the continuity equation with the
power-law source function $Q=K\gamma_e^{-p}$, normalized by a
local injection rate (Moderski, Sikora \& Bulik 2000). The cooling
of the electrons due to both synchrotron and inverse Compton has
been considered. The synchrotron radiation of the forward shock
electrons on the ``equal arriving surface" (on which the emission
reaches us at the same time) has been calculated strictly. The
synchrotron self-absorption has also been taken into account
strictly.

\begin{figure}
\begin{picture}(0,220)
\put(0,0){\includegraphics{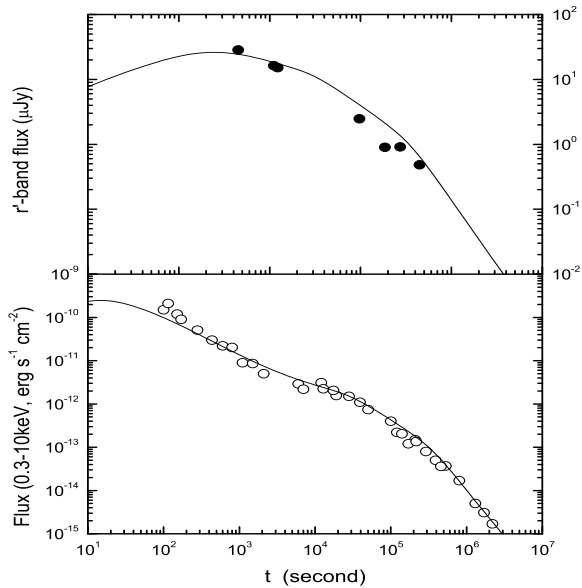}}
\end{picture}
\caption{Modeling the XRT and R-band afterglow light curves of GRB
051221A with energy injection from an underlying millisecond
magnetar. The optical and X-ray data are taken from Soderberg et
al. (2006) and Burrows et al. (2006), respectively. The $r'-$band
extinction of our Galaxy $\sim 0.19$ mag and that of the host
galaxy $\sim 0.2$ mag (Verkhodanov et al. 2000) have been taken
into account. The solid lines are our numerical results. Fitted
parameters of the GRB jet undergoing energy injection are
presented in Section 3.} \label{fig:051221A}
\end{figure}

We consider a uniform relativistic jet undergoing the energy injection from the
central source and sweeping up its surrounding uniform medium. The energy injection
has been taken as $dE/dt=2\times 10^{48}~{\rm erg~s^{-1}}(1+t/1.5\times10^4)^{-2}$
for $t<1.5\times 10^4$ s otherwise $dE/dt=0$ (i.e., we assume that the supermassive
magnetar collapses when it has lost significant part of its angular momentum). As
usual, the fractions of shock energy given to the electrons and the magnetic field
(i.e., $\epsilon_e$ and $\epsilon_B$) are assumed to be constant. Shown in Figure
\ref{fig:051221A} is our numerical fit in the X-ray and optical bands with the
following jet parameters: $E_k=10^{52}\,{\rm erg}$, $\epsilon_e=0.3$,
$\epsilon_B=0.0002$, the circumburst density $n=0.01\,{\rm cm^{-3}}$, the
half-opening angle $\theta_j=0.1$, and the viewing angle $\theta_{\rm obs}$=0 (i.e.
on-beam viewing).

In addition, we find that a reverse shock emission component,
besides the forward shock emission, should be evoked to account
for the radio data, which is consistent with the finding in
Soderberg et al. (2006). In the pulsar/magnetar energy injection
scenario, the reverse shock emission has been calculated in Dai
(2004), assuming that the outflow is electron/positron pairs
dominated and the reverse shock parameters are similar to, or even
larger than that of the forward shock. The resulting reverse shock
emission is so bright that long-time flat bump should be evident
in multi-band afterglow lightcurves. This prediction is not
confirmed by current observations. This puzzle, however, could be
resolved if the fraction of the reverse shock energy given to the
electrons is very small, as shown below.

In our numerical code, the reverse shock dynamics/emission has not
been taken into account. It is investigated analytically instead.
We assume that the reverse shock emission accounting for the radio
data is powered by the energy injection. Following Sari \& Piran
(1999), after the reverse shock crosses the ejecta at $t_\times$
(noted that at $t\sim t_\times \sim 0.2$ day the central magnetar
collapses), the observed reverse shock emission flux
$F_{\nu_\oplus}\propto (t/t_\times)^{-2}$ for
$\nu_m<\nu_\oplus<\nu_c$, where $\nu_m$ is the typical synchrotron
emission frequency of the shocked electrons and $\nu_c$ is the
cooling frequency. On the other hand, $F_{\nu_\oplus}\propto
(t/t_\times)^{-1/2}$ for $\nu_\oplus<\nu_m<\nu_c$. The current
radio observation thus requires a $\nu_m(t_\times)\leq 8.46$ GHz.
At $t_\times \sim 0.2$ day, the ejecta is at a radius $R \sim
7\times 10^{17}$ cm and moves with a bulk Lorentz factor
$\Gamma(t_\times) \sim 20$ (obtained in our numerical
calculation), thus the comoving toroidal magnetic field of the
magnetar wind is $B_w \sim [2L_{\rm dip}/(R^2 \Gamma_w^2
c)]^{1/2}\sim 20/\Gamma_w$ Gauss, where $\Gamma_w$ is the bulk
Lorentz factor of the magnetar wind. In the reverse shock phase,
the toroidal magnetic field will be amplified by a factor $\sim
\Gamma_{\rm rsh}\approx \Gamma_w/2\Gamma(t_\times)$ for
$\Gamma_w\gg \Gamma(t_\times)$ (Kennel \& Coronitti 1984). On the
other hand, the constraint $\nu_m(t_\times) \approx 2.8\times 10^6
{\rm Hz}~{\gamma_m}_{(t_\times)}^2 \Gamma(t_\times) \Gamma_{\rm
rsh} B_w /(1+z) \leq 8.46{\rm GHz}$ yields
${\gamma_m}_{(t_\times)} \leq 20$, where ${\gamma_m}_{(t_\times)}$
is the minim random Lorentz factor of the electrons accelerated in
the reverse shock front. Such a small\footnote{The numerical
coefficient $3/4$ is adopted because in the ideal MHD shock jump
condition, the random Lorentz factor of the post-shock protons is
proportional to $3(\Gamma_{\rm rsh}-1)/4$ rather than $\Gamma_{\rm
rsh}-1$ (see eq. (17) of Fan, Wei \& Zhang 2004b; see also Fan,
Wei \& Wang 2004a and Zhang \& Kobayashi 2005 for numerical
verification).} ${\gamma_m}_{(t_\times)}\equiv 3\epsilon_e
(p-2)(\Gamma_{\rm rsh}-1)m_p/[4(p-1)m_e]$ requires that the
fraction of shock energy given to the electrons $\epsilon_e$ is in
order of $m_e/m_p$ because it is very likely that now $\Gamma_w\gg
\Gamma(t_\times)$, where $m_e$ and $m_p$ are the rest mass of the
electron and the proton, repsectively. Taking $\epsilon_e \sim
m_e/m_p$ and $p\sim 2.3$, we thus have $\Gamma_{\rm rsh}\leq 115$
and $\Gamma_w \simeq 2\Gamma_{\rm rsh}\Gamma(t_\times) \leq 4600$.
After the reverse shock phase, the magnetic energy may be
translated to the forward shock mainly by the magnetic pressure
working on the initial GRB ejecta and the shocked medium. However,
the details are far from clear.

At $t\sim 0.91$ day, the 8.46 GHz emission flux is $0.155$ mJy.
Therefore the optical ($4.6\times 10^{14}$ Hz) flux at $t_\times
\sim 0.2$ day is $\leq 0.155{\rm mJy}(0.91/0.2)^2(4.6\times
10^{14}{\rm Hz}/8.46{\rm GHz})^{-0.65}=2.7~{\rm \mu Jy}$. It is
about one order lower than the forward shock optical emission. As
for the X-ray, now $\nu_c (t_\times)\approx 2\times 10^{15}$ Hz,
 the X-ray (at 2 keV) flux at $t_\times \sim 0.2$ day is $\leq 0.155{\rm
mJy}(0.91/0.2)^2(2\times 10^{15}{\rm Hz}/8.46{\rm
GHz})^{-0.65}(4.84\times 10^{17}/2\times 10^{15})^{-1.15}\sim
10^{-6}~{\rm mJy}$, is also significantly lower than the forward
shock X-ray emission. Our estimates made here are independent of
the poorly known magnetized reverse shock dynamics. We thus
conclude that the reverse shock X-ray/optical emission is
unimportant and can be ignored.

One caveat is that the dipole radiation of the magnetar is almost isotropic. The
medium surrounding the magnetar but out of the GRB ejecta cone would be accelerated
by the energetic wind. Because the energy injected into the GRB ejecta is larger
than that contained in the initial ejecta, the outflow contributing to the afterglow
emission would be close to be an isotropic fireball rather than a highly jetted
ejecta. So the magnetar energy injection model is somewhat challenged by the late
jet break detected in GRB 051221A. However, this puzzle could be resolved if in
other directions a large amount of baryons ($\sim 0.01~M_\odot$) ejected in the
double neutron star merger have existed there, as found in the previous numerical
simulations (Rosswog et al. 1999; Ruffert \& Janka 2001). The ejected material would
be accelerated by the magnetar wind to a bulk Lorentz factor $\sim $ a few, provided
that $E_{\rm inj}\sim 10^{52}$ erg. This wide but only mild-relativistic outflow
will give rise to a very late ($\sim 10^6-10^7$ s after the burst) multi-wavelength
re-brightening. However, for typical parameters taken in this Letter (the isotropic
energy is $1.5\times 10^{52}$ erg and the initial Lorentz factor is 5.0, i.e.,
assuming the material ejected from the merger is about $1.5\times 10^{-3}~M_\odot$),
the flux is not bright enough to be detectable. The emission peaks at $t\sim 3\times
10^6$ s. The 0.3-10 keV flux is $\sim 1\times 10^{-15}~{\rm erg~s^{-1}~cm^{-2}}$
which is marginal for the detection of {\it Chandra}, and the 8.46 GHz flux is $\sim
0.01$ mJy. Both are consistent with the extrapolation of current observations
(Soderberg et al. 2006; Burrows et al. 2006). If an event like GRB 051221A takes
place much closer, for example, at $z\sim 0.1$, such very late multi-wavelength
re-brightening may be detectable for {\it Swift} XRT or {\it Chandra} and other
radio telescopes. This predication could be tested in the coming months or years.

\section{Discussion and Summary}
Short hard GRBs may be powered by the merger of double neutron
stars (e.g. Eichler et al. 1989). For GRB 051221A, ruling out of
bright supernova component in the late optical afterglow
lightcurve suggests that the progenitor probably is not a massive
star and instead is consistent with the double neutron star merger
model. After the energetic merger, a black hole (Eichler et al.
1989), or a differentially rotating neutron star (Klu\'zniak \&
Ruderman 1998; Dai et al. 2006), or a magnetar may be formed.
Among the above cases the last one is based on various dynamo
mechanisms (Rossowog, Ramirez-ruiz \& Davis 2003; Gao \& Fan 2006
and references therein) and in particular the MHD simulation of
the two neutron star coalescence (Price \& Rosswog 2006). How long
can the supermassive magnetar survive? In the general-relativistic
numerical simulation, the resulting hypermassive magnetar
collapses in a very short time $\sim 100$ ms (Shibata et al.
2006). The hypermassive magnetar has a mass exceeding the mass
limit for uniform rotating but the supermassive magnetar does not.
Therefore a supermassive magnetar may be able to survive in a long
time as shown in Duez et al. (2006). In view of the uncertainties
involved in the numerical simulation, some constraints from the
observation rather than just from the theoretical calculation are
needed.

Thanks to the successful running of {\it Swift} and {\it Chandra},
now short GRBs could be localized rapidly and their X-ray
afterglows could be monitored continuously. The X-ray flat segment
in GRB 051221A strongly suggests a significant energy injection.
Such an energy injection could be well accounted for if the
central engine is a millisecond pulsar with a dipole magnetic
field $\sim 10^{14}$ Gauss. The X-ray flat segment in GRB 051221A
thus provides us a possible evidence for a long time living
magnetar formed in the double neutron star merger. In this
scenario, the material ejected from the merger would be swept and
accelerated by the strong magnetar wind. This wide but
mild-relativistic component would give rise to a very late
multi-wavelength re-brightening and might be detectable for an
event like GRB 051221A but much closer.

We would like to point out that the magnetar energy injection
model is not unique to account for the data. For example, assuming
the energy carried by the material of the initial GRB ejecta
satisfies the relation $E(>\Gamma)\propto \Gamma^{-4.5}$ (Rees \&
M\'esz\'aros 1998; Sari \& M\'esz\'aros 2000), the X-ray afterglow
lightcurve of GRB 051221A could also be well reproduced (Soderberg
et al. 2006; Burrows et al. 2006). However, the physical process
pumping such kind of energy injection, in particular in the short
GRB scenario, is not clear yet.

\section*{Acknowledgments}
We thank Tsvi Piran for an important remark and Jens Hjorth for
valuable comments that improved this paper. We also thank the
referee and X. W. Liu for constructive suggestions. YZF is
supported by the National Natural Science Foundation (grants
10225314 and 10233010) of China, the National 973 Project on
Fundamental Researches of China (NKBRSF G19990754), and US-Israel
BSF. DX is at the Dark Cosmology Centre funded by The Danish
National Research Foundation.

\end{document}